\documentclass[aps,prb,superscriptaddress,reprint,bibliography]{revtex4-1} 
\usepackage[section]{placeins}
\usepackage{soul}

\usepackage{lineno}
\usepackage{graphicx}  % needed for figures
\usepackage[caption=false]{subfig}
\usepackage{dcolumn}   % needed for some tables
\usepackage{tabularx}
\usepackage{bm}        % for math
\usepackage{amssymb}   % for math
\usepackage{amsmath,stmaryrd}
\usepackage{dsfont}
\usepackage{booktabs}
\usepackage{blkarray, multirow, graphicx, diagbox, color, colortbl}
\usepackage[dvipsnames]{xcolor}
\usepackage{bbm, bbold}
\usepackage{glossaries}
\usepackage{hyphenat}
\usepackage{ifthen}
\usepackage[colorlinks, linkcolor = blue, citecolor = blue, filecolor = black, urlcolor = blue]{hyperref}
\usepackage{xkeyval}
\usepackage{moreverb}
\usepackage{rotating}
\usepackage{wrapfig}
\usepackage{xspace}
\usepackage{nicefrac}
\usepackage[]{units}
\usepackage{physics}
\usepackage{braket}
\usepackage[inline]{enumitem}
\usepackage{tabto}
\usepackage{listings}
\usepackage{xstring}
\usepackage{pgfplots}
\usepackage[version=4]{mhchem}
\pgfplotsset{compat=1.18}
\def\ReplaceStr#1{%
	\IfSubStr{#1}{p}{%
		\StrSubstitute{#1}{p}{.}}{#1}}

\usepackage[capitalise]{cleveref} %
\usepackage{chemformula}
\usepackage{algorithm}

\newcommand{\arrowline}[2]{%
  \mathrel{%
    \begin{tikzpicture}[baseline={(0,0)}]
      \draw[thick, color=#1] (-0.3,0.09) -- (0.3,0.09);
      \ifthenelse{\equal{#2}{up}}{
        \draw[->, thick, >=stealth] (0,-0.1) -- (0,0.3);
      }{
        \draw[->, thick, >=stealth] (0,0.3) -- (0,-0.1);
      }
    \end{tikzpicture}%
  }%
}

\definecolor{colorA}{rgb} {0.9607843137254902, 0.25882352941176473, 0.25882352941176473} % red
\definecolor{colorB}{rgb} {0.9607843137254902, 0.6666666666666666, 0.25882352941176473} % orange
\definecolor{colorC}{rgb} {0.6, 0.6, 0.6} % grey
\definecolor{colorD}{rgb} {0.6549019607843137, 0.9607843137254902, 0.25882352941176473} % green
\definecolor{colorE}{rgb} {0.25882352941176473, 0.9607843137254902, 0.8784313725490196} % blue
\graphicspath{{figures/main_text/}}

\begin{document}

\author{Jiaqi Wu} 
\email{jw2311@cam.ac.uk}\affiliation{Theory of Condensed Matter Group, Cavendish Laboratory, University of Cambridge, J\,J\,Thomson Avenue, Cambridge CB3 0US, UK}

\author{Leonard Werner Pingen} 
\affiliation{Theory of Condensed Matter Group, Cavendish Laboratory, University of Cambridge, J\,J\,Thomson Avenue, Cambridge CB3 0US, UK}

\author{Asa Kelsey}
\affiliation{Theory of Condensed Matter Group, Cavendish Laboratory, University of Cambridge, J\,J\,Thomson Avenue, Cambridge CB3 0US, UK}
\affiliation{Queens' College, University of Cambridge, Silver Street, Cambridge CB3 9ET, UK}

\author{Han Yan}
\affiliation{Institute for Solid State Physics, The University of Tokyo, Kashiwa, Chiba 277-8581, Japan}

\author{Timothy K. Dickens} 
\affiliation{Peterhouse, University of Cambridge, Trumpington Street, Cambridge CB2 1RD, UK}

\author{Bo Peng} 
\email{bp432@cam.ac.uk}
\affiliation{Theory of Condensed Matter Group, Cavendish Laboratory, University of Cambridge, J\,J\,Thomson Avenue, Cambridge CB3 0US, UK}

\def\thetitle{Symmetry-induced magnetic fullerene}

\title{\thetitle}

\begin{abstract}
Defect-free, charge-neutral, pure-carbon materials are generally viewed as intrinsically non-magnetic. 
Here we challenge this view by establishing a fundamental principle of symmetry-induced magnetism in pure-carbon fullerene systems through molecular orbital theory. 
We show that high-order symmetry of fullerene molecules or their crystalline lattices induces degenerate energy levels and hence, quantised magnetic moments at half filling. This mechanism holds for \textit{all} high-order symmetries. 
It suggests a plausible intrinsic origin for the previously debated magnetic fullerene. 
Additionally, recent experimental advances in the synthesis of monolayer fullerene networks provide a feasible platform to implement our prediction on the deep link between symmetry and magnetism in these systems. 
Our results create new, broad frontiers of quantum magnetism by introducing molecular or crystalline symmetries in \textit{any} lattice.
\end{abstract}

\maketitle 

Quantum magnetism provides a central route to correlated and topological phases, including symmetry-broken magnetic order, symmetry-protected states, altermagnetism, topological orders, and quantum spin liquids. Defect-free and charge-neutral carbon systems, however, are not generally viewed as natural platforms for quantum magnetism because they are usually regarded as closed-shell and intrinsically non-magnetic. When magnetism does appear in carbon, it is typically linked to defects, edges, radicals, or other open-shell $\pi$ fragments that host unpaired $p$-electron spins~\cite{McConnell1963,Mataga1968,Clar1972,Rode2004,Son2006,Ugeda2010,Yazyev2010,Magda2014,Du2023,Dickens2024,Dickens2025}.
Although this route has enabled magnetism to arise, it remains difficult to realise such magnetic motifs in chemically stable and structurally controlled periodic materials, even in one-dimensional spin architectures~\cite{Mishra2021}. 

Recent synthesis of 2D covalent fullerene networks points to a distinct route~\cite{Hou2022,Meirzadeh2023}: generating unpaired electrons not from defects, but from symmetry-enforced orbital degeneracies. Because both the molecular symmetry of fullerene building blocks and the lattice symmetry of their assemblies can be engineered~\cite{Wu2025,Peng2022c}, these systems provide an ideal platform to test whether symmetry itself can generate magnetic moments in otherwise closed-shell carbon frameworks. In carbon magnetism, 
the effects of symmetry have been rarely addressed. 
Under the point group symmetry operations of the molecule, the molecular orbitals must conform to an irreducible representation (irrep) of the group of molecule~\cite{Mulliken1933}. 
For point groups with a symmetry element of order 3 or higher, molecular orbitals can transform as 2D irreps and therefore be degenerate by symmetry~\cite{Cotton1991} (point groups with only symmetry elements of order 1 or 2, where each symmetry operation is its own inverse, must be Abelian and hence only have 1D irreps~\cite{Dummit2004}, see Section~S1 in the Supplementary Information). When such degenerate orbitals are partially filled, exchange favours spin polarisation of the unpaired electrons, giving rise to non-vanishing magnetic moments~\cite{Dunlap1990}.  The group-theoretic statement is general: high-order symmetry elements can support multidimensional irreps, which induces magnetism when the corresponding degenerate levels are partially filled.  
  
Here, we establish a systematic approach to inducing magnetic moments in fullerene monolayers through high-order molecular or crystalline symmetries. We first demonstrate the universal mechanism in molecular systems that display order 3 and 4 symmetry elements. This symmetry-induced mechanism is predicted by molecular-orbital analysis and confirmed by first-principles calculations.  
We next examine systems where the high-order symmetry is absent on the molecular level but emerges from the lattice geometry, showing the robustness and versatility of symmetry-induced magnetism in the absence of high-order molecular symmetry. We further demonstrate that this platform can realise quantum anomalous Hall phases and unconventional local magnetic excitations. Finally, we discuss experimentally relevant fullerene systems in which doping can activate the same symmetry-based mechanism. 

\begin{figure*}
    \centering
    \includegraphics[width=\linewidth]{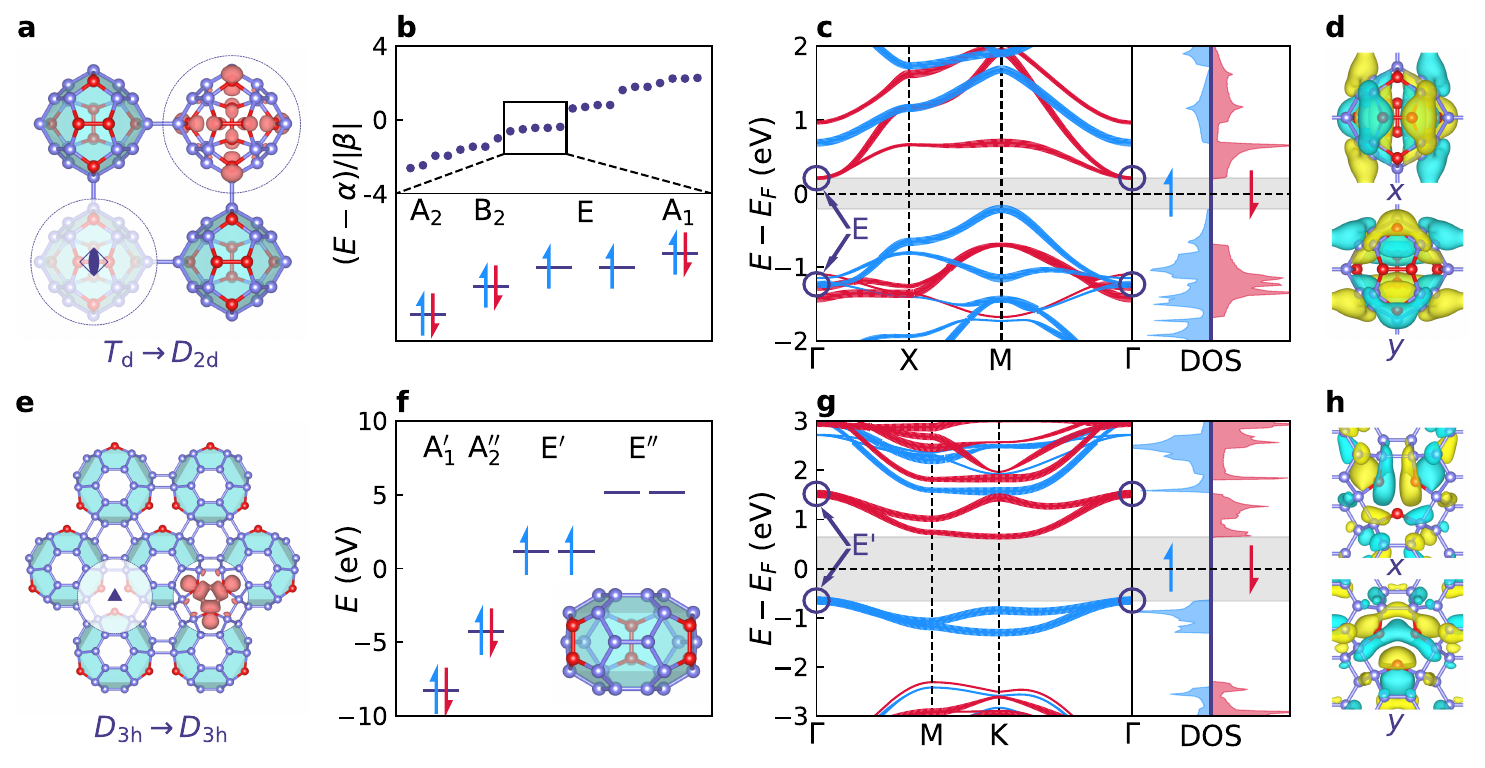}
    \caption{\textbf{Magnetism from high-order symmetry elements. } 
    \textbf{a} Top view of the crystal structure with spin densities in inset (isosurface levels at 0.012\,$a_0^{-3}$, where $a_0$ is the Bohr radius), \textbf{b} molecular orbitals, and \textbf{c} spin-polarised band structure with the density of states of square lattice C$_{28}$ monolayer with $S_4$ symmetry. The Kohn-Sham wavefunctions of the two degenerate bands at $\Gamma$ are shown in \textbf{d}. 
    \textbf{e} Top view of the crystal structure with spin densities in inset (isosurface levels at 0.003\,$a_0^{-3}$), \textbf{f} molecular orbitals, and \textbf{g} spin-polarised band structure with the density of states of triangular lattice C$_{36}$ monolayer with $C_3$ symmetry. The Kohn-Sham wavefunctions of the two degenerate bands at $\Gamma$ in the vicinity of the principal magnetic atoms are shown in \textbf{h}. 
    The high-order symmetry axis is marked on each crystal structure in \textbf{a} and \textbf{e}. A side view of the C$_{36}$ molecule is shown in the inset of \textbf{f}. 
    The linewidth of bands corresponds to the weight of the $p$ Wannier functions associated with the principal magnetic atoms, as marked in red in \textbf{a} and \textbf{e}. 
    }
    \label{fig:element}
\end{figure*} 

\vspace{6mm}\noindent\textbf{\large Results}\vspace{1mm}

\vspace{1.5mm}\noindent\textbf{Magnetism induced by molecular symmetry} \vspace{0.5mm}

To illustrate how molecular symmetry can induce magnetic moments, we first consider a fullerene monolayer whose molecular building blocks retain an $S_4$ symmetry axis, as shown in Fig.~\ref{fig:element}\textbf{a}.
An isolated C$_{28}$ molecule has the point group $T_\mathrm{d}$. When crystallised into a 2D square lattice with space group $P\bar{4}m2$, its molecular symmetry is reduced to $D_\mathrm{2d}$ with $S_4$ symmetry. In the monolayer, neighbouring C$_{28}$ molecules are connected by C--C single bonds along the $x$ and $y$ directions. First-principles calculations predict a quantised magnetic moment of $2.0\,\mu_{\mathrm{B}}$ per fullerene, with spin density delocalised over the eight red $sp^2$ carbon atoms in Fig.~\ref{fig:element}\textbf{a}; the resulting moments form an effective spin-1 square lattice.

To understand the symmetry origin of this molecular magnetism, we analyse the molecular-orbital spectrum of the fullerene building block. Because the electronic states near the Fermi level are dominated by $\pi$ electrons in fullerene monolayers~\cite{Wu2025}, we calculate the $\pi$-orbital energies using the H{\"u}ckel method~\cite{Hueckel1932,Kutzelnigg2007}, with one $p$ orbital assigned to each $sp^2$ carbon atom. Each $p$ orbital is assigned with an onsite energy $\alpha$, while nearest-neighbouring $p$ orbitals are coupled by a hopping amplitude $-\beta$. The calculated energies are plotted in Fig.~\ref{fig:element}\textbf{b}. For the 2D square C$_{28}$ lattice, the molecular point group $D_\mathrm{2d}$ contains a 2D irrep E associated with the principal four-fold improper rotation axis $S_4$ perpendicular to the monolayer plane. Counting down from the highest occupied molecular orbital (HOMO), the H{\"u}ckel spectrum contains five relevant molecular-orbital levels,
two of which form the degenerate E doublet. In the ground-state filling, this E doublet is half-filled, whereas the remaining three orbitals, which transform under 1D irreps, are fully occupied (for the relative energies of the E and A$_1$ orbitals from molecular-orbital theory, 
see Section~S2 in the Supplementary Information).
This identifies the $S_4$ principal axis as the symmetry element responsible for the half-filled E degeneracy that enables the quantised magnetic moment. 
We remark here that an alternative model for the origin of magnetism can also be constructed based on the conventional resonating valence bond picture for unpaired electrons~\cite{Peng2026}. However, the molecular-orbital description used here gives quantitative predictions for the relevant energy levels and provides a systematic design principle for magnetic carbon based on molecular and crystalline symmetry. 

Figure~\ref{fig:element}\textbf{c} shows the band structure of square lattice C$_{28}$ monolayer from unscreened hybrid functional calculations. Near the Fermi level, two bands remain degenerate at $\Gamma$, as expected from the E doublet. These two singly occupied bands transform under the E irrep, corresponding to the half-filled E orbitals predicted by the H{\"u}ckel analysis. In the \textit{ab initio} electronic structure, the ordering of the five frontier orbitals is modified by intermolecular interactions that are neglected in the H{\"u}ckel model; nevertheless, their wavefunctions are well captured by the molecular-orbital analysis (see Section~S2 in the Supplementary Information). Thus, the magnetism associated with the singly occupied E-degenerate bands can be traced to the principal $S_4$ axis of the molecular point group.
Figure~\ref{fig:element}\textbf{d} shows the Kohn-Sham wavefunctions of the two degenerate E bands at $\Gamma$, labelled according to their corresponding Cartesian functions. 
This confirms that the E degeneracy reflects the symmetry equivalence of the $x$ and $y$ directions enforced by the principal $S_4$ axis.

\vspace{1.5mm}\noindent\textbf{Magnetism induced by crystalline symmetry}\vspace{0.5mm}

Having established how molecular symmetry induces magnetism in square lattice C$_{28}$ monolayer, we next consider a case in which the relevant degeneracy is controlled not by the isolated fullerene unit, but by the local crystalline environment. The 2D triangular lattice of C$_{36}$ is another magnetic fullerene monolayer, with space group $P\bar{6}m2$, as shown in Fig.~\ref{fig:element}\textbf{e}. Each fullerene unit is surrounded by six neighbouring molecules, with adjacent fullerenes connected by three intermolecular cycloaddition bonds. An isolated C$_{36}$ molecule has the dihedral point group $D_\mathrm{3h}$. In a closely packed triangular lattice, the local symmetry also retains $D_\mathrm{3h}$ symmetry, which supports 2D irreps E$^\prime$ and E$^{\prime\prime}$. Although the relevant point group differs from that of square lattice C$_{28}$ monolayers, monolayer triangular lattice C$_{36}$ also possesses a quantised magnetisation of $2.0\,\mu_\mathrm{B}$ per fullerene, forming an effective spin-1 triangular lattice. The spin density is concentrated on the six red atoms, which form three vertical $\sigma$-bonded pairs near three interstices around each fullerene (see inset of Fig.\,\ref{fig:element}\textbf{f}).

A molecular resonating-valence-bond picture for an isolated C$_{36}$ unit predicts neither an unpaired electron nor a local moment: the eighteen $sp^3$ carbon atoms are fully saturated; the twelve $sp^2$ atoms forming two six-membered rings on the top and bottom surfaces are isoelectronic to benzene and hence closed shell; and each isolated pair of red atoms would be expected to form a double bond without spin polarisation. To explain the magnetic moment that appears in the crystal, 
we construct molecular orbitals for the three closely packed pairs of magnetic atoms marked by red colour around half of the interstices in Fig.~\ref{fig:element}\textbf{e}.
The local point group of each such interstitial environment is $D_\mathrm{3h}$ and therefore supports 2D irreps. The pairs at the ends of the red $\sigma$ bonds overlap weakly sideways in a $\pi$-bond-like manner. We therefore extract the hopping energies between the relevant $p$-type Wannier functions from \textit{ab initio} calculations (see Section~S2 in the Supplementary Information). As shown in Fig.~\ref{fig:element}\textbf{f}, the resulting tight-binding energy levels demonstrate that the degenerate E$^\prime$ HOMO hosts two unpaired electrons. Thus, the magnetism in triangular lattice C$_{36}$ is more naturally traced to the symmetry of the local interstitial environment than to that of an isolated fullerene building block. 

Figure~\ref{fig:element}\textbf{g} shows the \textit{ab initio} electronic structure of triangular lattice C$_{36}$ calculated using an unscreened hybrid functional. Two singly occupied bands appear at the valence-band maximum (VBM) and are isolated from the lower valence bands. These singly occupied bands are dominated by contributions from the six magnetic atoms. The band crossing at $\Gamma$ transforms as the 2D irrep E$^\prime$, consistent with the predicted molecular orbitals. The two corresponding wavefunctions in Fig.~\ref{fig:element}\textbf{h} follow the well-known coefficients of $(0,1,-1)$ and $(2,-1,-1)$ for the three-fold rotation symmetry. 

\begin{figure}
    \centering
    \includegraphics[width=\linewidth]{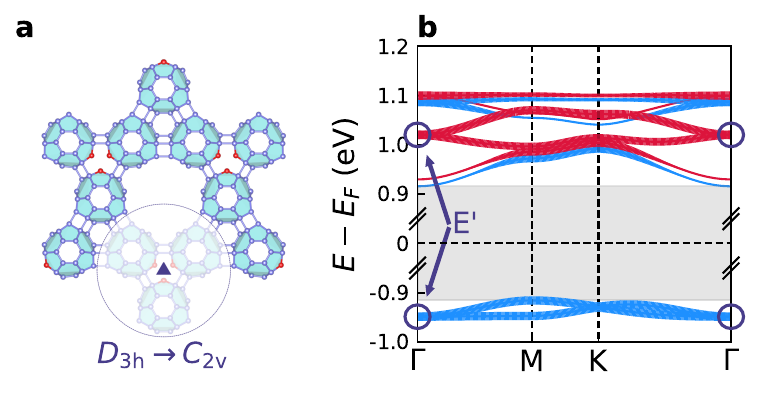}
    \caption{\textbf{Magnetism without molecular symmetry.}
    \textbf{a} Top view of the crystal structure and \textbf{b} band structure of Kagome lattice C$_{36}$. The linewidth of bands corresponds to the weight of the $p$ Wannier functions associated with the principal magnetic atoms, as marked in red in \textbf{a}.
    }
    \label{fig:kagome}
\end{figure}

\vspace{1.5mm}\noindent\textbf{A general mechanism for high-order symmetries}\vspace{0.5mm}

Having established that symmetry-induced magnetism need not originate from the symmetry of an isolated fullerene molecule, we now examine a case in which the fullerene building block lacks high-order molecular symmetry, while the crystalline environment retains it. Figure~\ref{fig:kagome}\textbf{a} shows the crystal structure for a Kagome lattice of C$_{36}$ with space group $P\bar{6}m2$, which can be viewed structurally as a vacancy-ordered derivative of triangular C$_{36}$ lattice. The point group of each fullerene unit is reduced from $D_\mathrm{3h}$ to $C_\mathrm{2v}$, which has only 1D irreps and therefore cannot by itself enforce a degenerate molecular orbital. However, the local interstitial orbital fragment retains $D_\mathrm{3h}$ symmetry, allowing its orbitals to transform as 2D irreps and hence to form spin-polarised degenerate bands at $\Gamma$. Figure~\ref{fig:kagome}\textbf{b} shows the corresponding band structure of Kagome lattice C$_{36}$ monolayer. The degenerate bands at $\Gamma$ have strong weight on the principal magnetic atoms marked in red in Fig.~\ref{fig:kagome}\textbf{a}, indicating that the magnetic moment is localised around the interstitial fragments. This provides a clean proof of principle that high-order molecular symmetry is not required for symmetry-induced magnetism: the relevant degeneracy can instead be induced by the symmetry of the local crystalline environment. 

\begin{figure*}
    \centering
    \includegraphics[width=\linewidth]{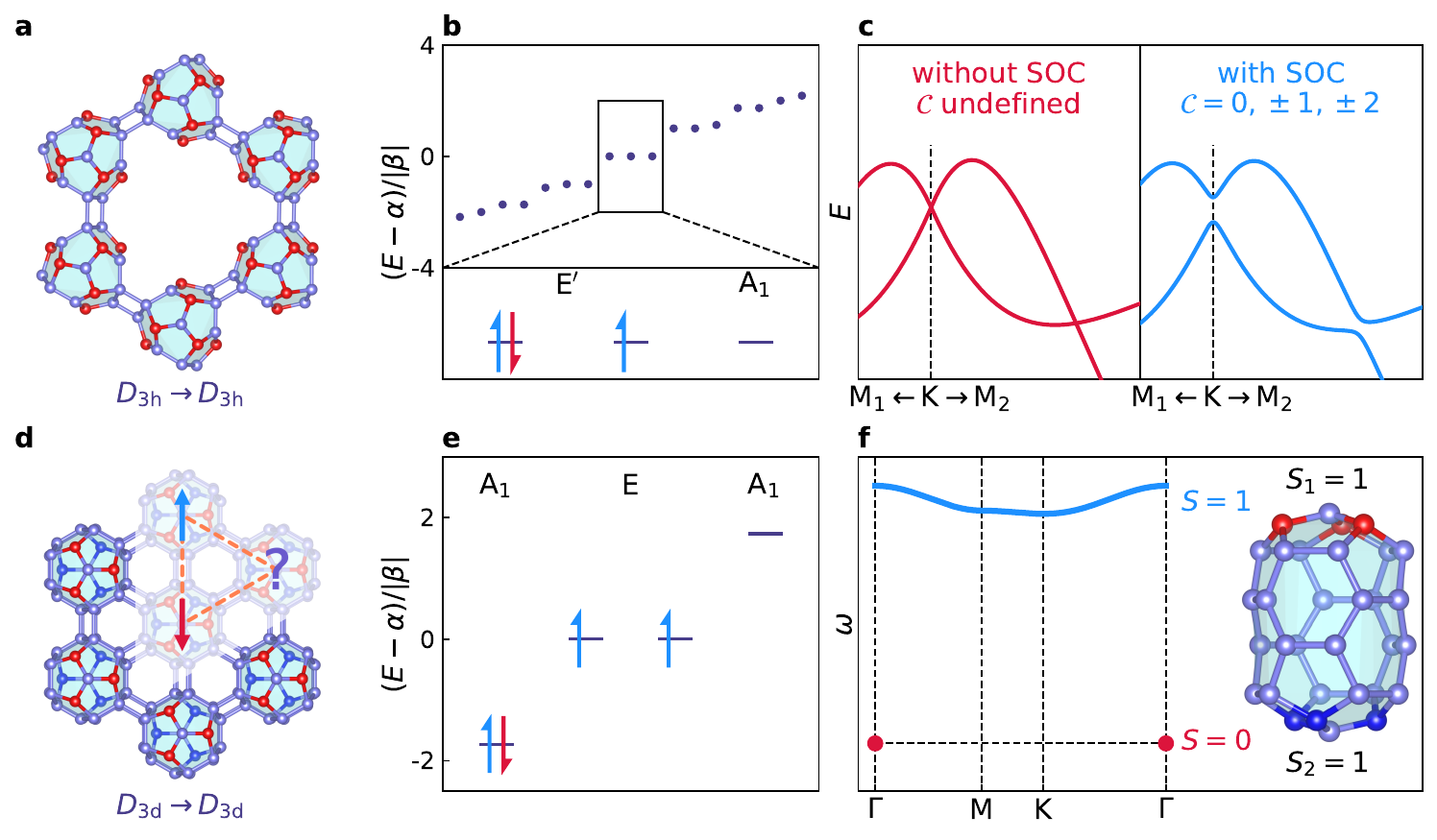}
    \caption{\textbf{Fullerene quantum materials.} \textbf{a} Top view of the crystal structure, \textbf{b} molecular orbitals, and \textbf{c} band structure with and without spin-orbit coupling of monolayer honeycomb C$_{26}$, showing band inversion and opening of a direct band gap. \textbf{d} Top view of the crystal structure of monolayer triangular lattice C$_{32}$, \textbf{e} molecular orbitals of the trimethylenemethane-like fragments, and \textbf{f} schematic dispersion of triplon excitations.}
    \label{fig:application}
\end{figure*}

\begin{figure*}
    \centering
    \includegraphics[width=\linewidth]{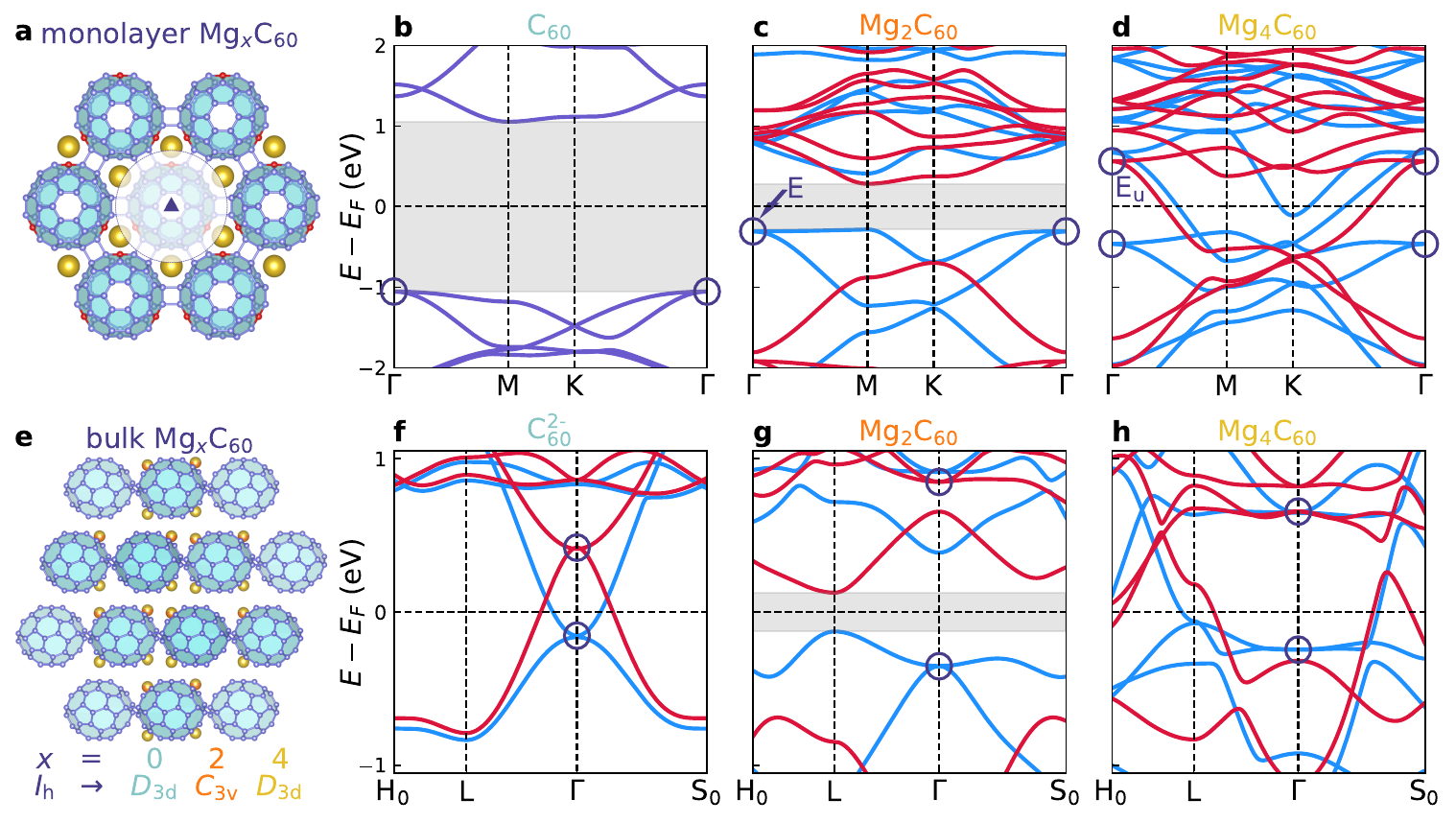}
    \caption{\textbf{Magnetism in experimentally synthesised C$_{60}$ systems.} \textbf{a} Top view of the crystal structure of monolayer r-triangular lattice Mg$_x$C$_{60}$. 
    Band structures of \textbf{b} non-magnetic pristine r-triangular lattice C$_{60}$, \textbf{c} Mg$_2$C$_{60}$ and \textbf{d} Mg$_4$C$_{60}$ monolayers. 
    \textbf{e} Side view of the crystal structure of bulk rhombohedral Mg$_x$C$_{60}$. 
    Band structures of bulk rhombohedral \textbf{f} $2e^-$-doped C$_{60}$, \textbf{g} Mg$_2$C$_{60}$ and \textbf{h} Mg$_4$C$_{60}$. }
    \label{fig:R-3m}
\end{figure*}

The origin of magnetism in these fullerene monolayers can be rationalised from both a localised molecular or a delocalised crystal viewpoint. Locally, two half-filled degenerate molecular orbitals that transform under a 2D irrep result in two unpaired electrons in the triplet state, which makes each fullerene molecule a spin-1 unit. On a delocalised basis, the two bands formed by the degenerate orbitals at $\Gamma$ cannot open a band gap by pushing one band above the Fermi level due to the symmetry-enforced crossing. 
The two bands that are degenerate at $\Gamma$ result in pronounced peaks in the density of states near the Fermi energy in Fig.~\ref{fig:element}\textbf{c} and \textbf{g}, making the spin polarisation energetically more favourable through the Stoner mechanism. 

The molecular-orbital splittings associated with the frontier electronic states are typically several eV ($\sim10^4$~K), which are much larger than the energy scales of any experimentally relevant magnetic, structural, or thermal effects.
As a result, the symmetry-induced orbital degeneracies lead to robust magnetism, where the low-energy physics is governed by the filling, exchange interactions, and intermolecular couplings within this well-separated orbital manifold.

A natural question is whether the half-filled symmetry-induced degenerate orbitals are susceptible to Jahn--Teller distortions~\cite{JahnTeller1937,Bersuker2006JahnTeller}, which tend to lift the degeneracy and quench the magnetic moment. In principle, such distortions are expected whenever degenerate electronic orbitals are partially occupied. However, unlike isolated molecules~\cite{ChanceyOBrien1997C60JT,Gunnarsson1997Fullerides}, our systems are extended fullerene networks in which any symmetry-lowering distortion must compete with the elastic cost of coherent lattice deformations. In Section~S3 of the Supplementary Information, we present the phonon dispersions of two representative systems to demonstrate such competitions. 
 
The diverse molecular and crystalline point-group symmetries enable a vast design space for symmetry-induced magnetic fullerene systems. For periodic crystalline lattices, the crystallographic restriction theorem allows only rotational symmetry elements of order 2, 3, 4, or 6~\cite{Kittel1976}. 
We exploit $C_3$ and $S_4$ principal axes for magnetic fullerene monolayers in the main text, and additional examples based on $C_6$ and $S_6$ principal axes are provided in Section~S4 of the Supplementary Information, covering all high-order symmetry elements allowed in periodic fullerene monolayers (conventional fullerene molecules generally lack $C_4$ symmetry because this requires four-membered rings). Together, these examples show that the same symmetry--degeneracy--filling mechanism applies across all high-order symmetry elements available to periodic fullerene monolayers.

\vspace{1.5mm}\noindent\textbf{Fullerene as a platform for quantum materials}\vspace{0.5mm}

We next show that symmetry-induced magnetic fullerene networks provide a versatile platform for engineering exotic quantum phases. The same symmetry mechanism enables effective spin lattices, topological phases, and unconventional magnetic excitations. The intermolecular exchange interactions in magnetic fullerene monolayers have an energy scale of $0.1-30$\,meV depending on the overlap of wavefunctions between neighbouring fullerene molecules~\cite{Pingen2025,Wu2025b, Peng2026}. As an example, we find the energy difference between the ferromagnetic and N\'{e}el phases of square lattice C$_{28}$ monolayer to be 25\,meV per fullerene, suggesting a critical temperature of the order of $\sim300$\,K for an effective spin-1 square lattice. 

One immediate consequence of symmetry-induced ferromagnetism is the spontaneous breaking of time-reversal symmetry, which can be combined with spin-orbit coupling to realise topological phases. Figure~\ref{fig:application}\textbf{a} shows the crystal structure of monolayer honeycomb C$_{26}$. C$_{26}$ molecules exhibit the point group $D_\mathrm{3h}$, and the corresponding hexagonal lattice maintains this symmetry, featuring a $C_3$ axis. For each molecule, two orbitals transform under the E$^\prime$ irrep near the Fermi level, occupied by three electrons, producing a magnetic moment of $1\,\mu_\mathrm{B}$ per molecule in Fig.~\ref{fig:application}\textbf{b}. The resulting electronic structure resembles that of ferromagnetic graphene with broken time-reversal symmetry, providing a natural setting for quantum anomalous Hall effect. When spin-orbit coupling is included, a gap opens at the Dirac point at K in Fig.~\ref{fig:application}\textbf{c}, producing tuneable quantum anomalous Hall insulators with Chern numbers of $\pm2$, $\pm1$, and $0$ depending on the magnetisation direction~\cite{Pingen2025}.

Beyond single-particle topological phases, magnetic fullerene networks also provide a platform for interacting quantum spin models. As shown in Fig.~\ref{fig:application}\textbf{d}, in triangular lattice C$_{32}$, the molecular point group is $D_\mathrm{3d}$ with a principal $C_3$ axis. Monolayer triangular lattice C$_{32}$ has four $sp^2$ atoms on the top and bottom surfaces of the fullerene, forming a $\pi$ system on each surface analogous to trimethylenemethane~\cite{Dowd1966,Yarkony1974,Auster1982,Wenthold1996,Klein2001}. Within the H{\"u}ckel-theory description shown in Fig.~\ref{fig:application}\textbf{e}, each surface of the molecule hosts one spin-1 unit, arising from a singly occupied E orbital. On each surface of the monolayer, the intermolecular exchange interactions are antiferromagnetic, producing triangular networks of interacting spin-1 degrees of freedom as illustrated in Fig.~\ref{fig:application}\textbf{d}. The ground state is well approximated by a product of intramolecular $S=0$ singlets. The elementary magnetic excitations above this state carry quantum number $S=1$. These excitations are not conventional magnons arising from long-range magnetic order, but are instead triplon $S=1$ excitations in Fig.~\ref{fig:application}\textbf{f}. In the presence of spin-orbit coupling, these excitations can form three-level Dirac cones. Together, these examples illustrate that symmetry-induced magnetism in fullerene networks offers a unified route to both topological phases and strongly correlated quantum spin phenomena.

\vspace{1.5mm}\noindent\textbf{Experimental realisation}\vspace{0.5mm}

Finally, we discuss experimentally relevant routes to realising symmetry-induced magnetic fullerene. Monolayer triangular lattice and few-layer square lattice have already been synthesised for C$_{60}$ structures~\cite{Hou2022,Meirzadeh2023}.  
We therefore focus on C$_{60}$-based networks.
We start from a triangular lattice C$_{60}$ monolayer with $C_3$ symmetry in the space group $P\bar{3}m1$. The $C_3$-symmetric C$_{60}$ monolayer considered here, shown in Fig.~\ref{fig:R-3m}\textbf{a}, differs slightly from the experimentally reported triangular lattice structure~\cite{Hou2022} by a rotation of the C$_{60}$ molecules. We therefore denote it as r-triangular lattice C$_{60}$ to distinguish this structural variant from the reported triangular lattice phase. 

Figure~\ref{fig:R-3m}\textbf{b} shows the band structure of the r-triangular lattice C$_{60}$ monolayer. Although the pristine system is non-magnetic, the VBM at $\Gamma$ is doubly degenerate, consistent with the symmetry-based mechanism. This suggests that charge doping can move the Fermi level into partially filled degenerate orbitals and thereby induce magnetism. 

Here we focus on Mg-intercalated C$_{60}$ monolayers,
because it coincides exactly with the synthetic procedure~\cite{Tanaka2018,Hou2022,Argaman2023,Barnes2025}. Additionally, each Mg atom provides two electrons to the doubly generate bands to make them half-filled, as confirmed by the Bader charge analysis (see Section~S5 in the Supplementary Information). 
As shown in Fig.~\ref{fig:R-3m}\textbf{a}, 
the Mg atoms can be placed on one side of the monolayer, giving rise to Mg$_2$C$_{60}$ with space group $P3m1$; or on both sides, giving rise to Mg$_4$C$_{60}$ with space group $P\bar{3}m1$. Crucially, r-triangular lattice C$_{60}$, Mg$_2$C$_{60}$, and Mg$_4$C$_{60}$ have point groups $D_\mathrm{3d}$, $C_\mathrm{3v}$, and $D_\mathrm{3d}$, respectively, containing a $S_6$ or $C_3$ axis that enforces degenerate bands at $\Gamma$. 
Figure~\ref{fig:R-3m}\textbf{c} and \textbf{d} show the band structures of r-triangular lattice Mg$_2$C$_{60}$ and Mg$_4$C$_{60}$, respectively. Both systems exhibit spin polarisation, with magnetic moments of $2.0$ and $2.2\,\mu_\mathrm{B}$ per fullerene for Mg$_2$C$_{60}$ and Mg$_4$C$_{60}$, respectively. The Mg atoms transfer charge to the carbon network, tuning the Fermi level so that the degenerate orbitals become partially filled and host unpaired electrons. 

The same symmetry-based perspective also sheds light on rhombohedral C$_{60}$ van der Waals crystals, which have a long history in the discussion of carbon magnetism. Finite magnetisation was previously reported in nominally pure-carbon samples~\cite{Makarova2001a}, 
although subsequent studies attributed the observed magnetism to iron impurities~\cite{Han2003,Makarova2006,Talyzin2007}. Independently of this historical controversy, the rhombohedral space group contains high-order symmetry elements. The symmetry-based mechanism therefore predicts degenerate bands at $\Gamma$ in rhombohedral C$_{60}$ crystals, consistent with previous first-principles calculations~\cite{Kayley2025}. Similar to the r-triangular lattice C$_{60}$ monolayer, bulk rhombohedral C$_{60}$ is insulating in the neutral state, but charge doping can place the Fermi level in the symmetry-enforced degenerate orbitals. We consider three charge-tuning scenarios: electrostatic doping ($2e^{-}$-doped), one-sided Mg doping (Mg$_2$C$_{60}$), and two-sided Mg doping (Mg$_4$C$_{60}$), as shown in Fig.~\ref{fig:R-3m}\textbf{e}. These structures have space groups $R\bar{3}m$, $R3m$, and $R\bar{3}m$, respectively. Related charge-doped and Mg-intercalated fullerene systems have been realised experimentally~\cite{Iwasa1994,Nunez-Regueiro1995,Tanaka2018}. 
Figure~\ref{fig:R-3m}\textbf{f--h} shows the calculated band structures of $2e^{-}$-doped rhombohedral C$_{60}$, Mg$_2$C$_{60}$, and Mg$_4$C$_{60}$, respectively. All three cases show finite magnetisation once the symmetry-enforced degenerate states near $\Gamma$ are brought to the Fermi level and partially filled. These results do not resolve the historical debate of the reported magnetic fullerene~\cite{Makarova2001a,Han2003,Makarova2006,Talyzin2007}, but they identify a plausible intrinsic route to magnetism in symmetry-compatible fullerene crystals without invoking magnetic impurities.

To conclude, the calculated systems in Fig.~\ref{fig:element}--\ref{fig:R-3m} generally favour spin-polarised states over non-magnetic states, with the exception of one-sided doped r-triangular lattice Mg$_2$C$_{60}$ (see Section~S5 in the Supplementary Information). These results show that symmetry-induced magnetism can occur in undoped fullerene networks and can also be activated by charge tuning in experimentally relevant fullerene-based systems.   Additionally, various 2D lattices of multiple fullerene molecules (including C$_{36}$ and C$_{60}$) have been realised experimentally since three decades ago~\cite{Piskoti1998,Heath1998,Zhang2022a,Ding2023,Alotaibi2026}. Symmetry therefore provides an experimentally accessible design principle for realising quantum magnetism in carbon-based materials.

\vspace{6mm}\noindent\textbf{\large Discussion}\vspace{1mm}

We establish a general symmetry-based route to the realisation of magnetic fullerene: high-order molecular or crystalline symmetry can enforce degenerate electronic states, which generate local magnetic moments when partially filled. The essential ingredients are: (1) a high-order symmetry element that enforces a multidimensional irrep, (2) a filling that leaves the degenerate orbitals partially occupied, and (3) an exchange-driven spin-polarised state that is energetically competitive with non-magnetic metallic states or symmetry-lowered alternatives. We demonstrate this mechanism in defect-free, charge-neutral fullerene networks and further show that related C$_{60}$-based systems can be activated by charge tuning. Molecular-orbital and local-fragment-orbital analyses clarify whether the relevant degeneracy originates from the fullerene molecule itself or from the local crystalline/interstitial environment. This provides an analytically tractable description of the frontier electronic structure and a practical design rule for identifying magnetic fullerene candidates. More broadly, the same symmetry--degeneracy--filling principle should be applicable to other molecular networks, including covalent- and metal-organic frameworks, where molecular units and crystalline environments can be designed to support symmetry-enforced degeneracies.

\vspace{6mm}\noindent\textbf{\large Methods}\vspace{1mm}

Density functional theory (DFT)~\cite{Hohenberg1964,Kohn1965} calculations are performed using the Vienna \textit{ab initio} Simulation Package (\textsc{VASP})~\cite{Kresse1996,Kresse1996a} with the projector augmented-wave (PAW) basis set~\cite{Bloechl1994,Kresse1999} under the generalised gradient approximation (GGA) formalism. The Perdew-Burke-Ernzerhof functional revised for solids (PBEsol)~\cite{Perdew2008} is used for geometry optimisation. In cases where PBEsol underestimates the magnetisation (e.g., predicting a half-metal), the crystal structures are further relaxed using a hybrid functional of PBEsol mixed with $25\%$ unscreened Hartree-Fock exchange (denoted as PBEsol0)~\cite{Adamo1999}. The Brillouin zone is sampled using a $\Gamma$-centred $\mathrm{k}$-mesh of spacing $0.03\,\mathrm{\AA}^{-1}$, and the plane-wave cutoff is set as $800\,\mathrm{eV}$, which have been found to be converged parameters for fullerene systems~\cite{Wu2025,Peng2022c,Peng2023,Shaikh2025,Jones2023,Peng2025c,Peng2025a,Shearsby2025,Kayley2025} (see Section~S6 in the Supplementary Information for convergence tests). The interlayer vacuum spacing between monolayers is larger than $14\,\mathrm{\AA}$, and dipole corrections are used to eliminate interactions between periodic images~\cite{Makov1995}. Self-consistent field calculations are converged to $10^{-6}$\,eV, and the crystal structures are fully relaxed until the Hellmann-Feynman forces are less than $10^{-2}\,\mathrm{eV}\,\mathrm{\AA}^{-1}$. The crystal structures and spin densities are visualised by \textsc{VESTA}~\cite{vesta}. The electronic structures are calculated using the unscreened hybrid functional PBEsol0, as it gives more accurate descriptions of the band gaps of fullerene monolayers~\cite{Peng2022c} comparable to many-body perturbation theory~\cite{Champagne2024}. The Wannier functions~\cite{Marzari1997,Souza2001,Marzari2012} are constructed from all $\sigma$ bond centres, as well as from the additional $p$ orbital for $sp^2$ carbon atoms, as implemented in \textsc{Wannier90}~\cite{Mostofi2008,Mostofi2014,Pizzi2020}. The spin-polarised density of states is obtained from the Wannier-interpolated model on a $\mathbf{k}$-mesh of $200\times200\times1$, which provides comparable data with the tetrahedron method using Bl{\"o}chl corrections~\cite{Bloechl1994a}. The onsite exchange energies of the principal magnetic atoms are calculated as the difference between the corresponding spin-up and spin-down $p$ Wannier functions. Bader charge calculations are performed using the \textsc{BADER} program~\cite{Henkelman2006,Sanville2007,Tang2009}.

\vspace{6mm}\noindent\textbf{\large Acknowledgements}\vspace{1mm}

We thank Jiarui Cai and Rundong Yuan from the University of Cambridge for helpful discussions. J.W. acknowledges support from the Jardine Foundation, the Krishnan-Ang Foundation, and Trinity College Cambridge for a Jardine Trinity Scholarship. L.W.P. acknowledges support from the Klaus Höchstetter\hyp Stiftung at Munich and from the Trinity College Cambridge Studienstiftung des deutschen Volkes exchange studentship. 
H.Y. acknowledges the JSPS KAKENHI Grant No.~26K17090, Grant-in-Aid for Early-Career Scientists.
B.P. acknowledges support from Magdalene College Cambridge for a Nevile Research Fellowship. The calculations were performed using resources provided by the Cambridge Service for Data Driven Discovery (CSD3) operated by the University of Cambridge Research Computing Service (\url{www.csd3.cam.ac.uk}), provided by Dell EMC and Intel using Tier-2 funding from the Engineering and Physical Sciences Research Council (capital grant EP/T022159/1), and DiRAC funding from the Science and Technology Facilities Council (\url{http://www.dirac.ac.uk}), as well as with computational support from the UK Materials and Molecular Modelling Hub, which is partially funded by EPSRC (EP/T022213/1, EP/W032260/1 and EP/P020194/1), for which access was obtained via the UKCP consortium and funded by EPSRC grant ref EP/P022561/1.

\end{document}